\RequirePackage{fix-cm}

\documentclass[smallextended]{svjour3}       
\def\makeheadbox{{%
\hbox to0pt{\vbox{\baselineskip=10dd\hrule\hbox
to\hsize{\vrule\kern3pt\vbox{\kern3pt

\hbox{This is a post-peer-review, pre-copyedit version of this article.}
\hbox{The final authenticated version is available online at: [Insert DOI here].}
\kern3pt}\hfil\kern3pt\vrule}\hrule}%
\hss}}}
\smartqed  
\usepackage{graphicx}
\usepackage[export]{adjustbox}
\usepackage{wrapfig}
\usepackage{mathptmx}
\usepackage{subcaption}
\usepackage{soul}
\usepackage[dvipsnames]{xcolor}
\usepackage{latexsym}
\usepackage[utf8]{inputenc}
\usepackage[english]{babel}
\usepackage[numbers,sort&compress]{natbib}

\usepackage{libertine} 
\usepackage{siunitx}

\begin{document}

\title{Cryogenic amplification of image-charge detection for readout of quantum states of electrons on liquid helium 
}


\author{Asem Elarabi\textsuperscript{1}        \and
        Erika Kawakami\textsuperscript{1,2} \and Denis Konstantinov\textsuperscript{1} 
}

\authorrunning{Short form of author list} 
\institute{Asem Elarabi \at
\email{asem.elarabi@oist.jp}       
  \\
           \and
            Denis Konstantinov \at
           \email{denis@oist.jp}
           \and
         1 Quantum Dynamics Unit, Okinawa Institute of Science and Technology, Tancha 1919-1, Okinawa 904-0495, Japan \\
         2 PRESTO, Japan Science and Technology (JST), Kawaguchi, Saitama 332-0012, Japan
}


\maketitle

\begin{abstract}
Accurate detection of quantum states is a vital step in the development of quantum computing. Image-charge detection of quantum states of electrons on liquid helium can potentially be used for the readout of a single-electron qubit; however, low sensitivity due to added noise hinders its usage in high fidelity and bandwidth (BW) applications. One method to improve the readout accuracy and bandwidth is to use cryogenic amplifications near the signal source to minimize the effects of stray capacitance. We experimentally demonstrate a two-stage amplification scheme with a low power dissipation of \SI{90}{\micro\watt} at the first stage located at the still plate of the dilution refrigerator, and a high gain of \SI{40}{\decibel} at the second stage located at the 4 K plate. The good impedance matching between different stages and output devices ensure high BW and constant gain in a wide frequency range. The detected image-charge signals are compared for one-stage and two-stage amplification schemes.
\keywords{Cryogenic amplifier \and Electrons on Helium \and Qubits readout \and Heterojunction bipolar transistor \and Low temperature electronics}
\end{abstract}

\section{Introduction}
\label{intro}
Surface electrons (SE) on liquid helium is a promising platform for building scalable quantum computers \cite{Andrei1997,Dykman2003PRB,Platzman1999a}. Some progress has been made recently toward using this platform to realize qubits, in particular by utilizing the Rydberg states of electrons~\cite{Kawakami2019}. Furthermore, it was shown that the spin-state of electrons, which was also proposed as a basis for qubits \cite{Lyon2006a,Schuster2010}, potentially can be accessed by utilizing the spin-orbit interaction~\cite{Kawakami2019}. To achieve the quantum state readout for a single electron, a significant increase in the sensitivity and bandwidth (BW) of the detection method is vital. Many methods were proposed to realize higher sensitivity and signal-to-noise ratio (SNR) in quantum devices, one of which is by using low-noise amplifiers. However, multiple issues might limit the usage of amplifiers in detecting small signals from quantum devices residing at the mixing chamber of a dilution refrigerator.  Among these issues is the parasitic capacitance of the wires, which limits the bandwidth of the measured signal.
Additionally, the environmental noise and the active noise caused by the amplifier circuit deem small signals undetectable. To reduce these undesirable effects, it is preferable to use cryogenic amplifiers located as close to the qubits as possible \cite{ashoori1982}. In this way, the shortened wires result in less parasitic capacitance and environmental noise, in addition to the active noise reduction caused by cooling the amplifiers.

Josephson parametric amplifiers \cite{Stehlik2015,Abdo2014,Vijay2009}, SQUID amplifiers \cite{Schupp2020,Phipps2016,Korolev2014,Gremion2008,Korolev2011}, Single-electron transistor (SET) amplifiers \cite{Zimmerli1992,Devoret2000,Segall2002,Satoh1999}, and many other technologies were recently used for cryogenic small-signal low-noise amplification. However, most of these technologies require special driving devices and supporting optimized custom circuits placed at multiple locations inside the refrigerator, which adds to the complexity and increases costs. 
A more straightforward approach is to use a low power single transistor circuit placed near the qubit. Two types of commercially-available low-power transistors were proposed for cryogenic applications: high electron mobility transistor (HEMT) \cite{Tracy2018} and silicon-germanium (SiGe) heterojunction bipolar transistor (HBT) \cite{Curry2015,Wan2017,Curry2019,Zavjalov2019,Ivanov2016,Ivanov2011a,Weinreb2007,Goryachev2010}.  HEMT amplifiers are more suitable for high-frequency applications because of the reduced sensitivity at frequencies $< \SI{100}{\mega\hertz}$ caused by the large corner-frequency of the flicker noise \cite{Oukhanski2003,Angelov2000}. Furthermore, the power dissipation for HEMT-based amplifiers is larger than that for HBT-based amplifiers. 
On the other hand, HBT-based amplifiers are recognized to work at millikelvin temperature range with power dissipation as low as \SI{5}{\nano\watt} \cite{Curry2015}. Moreover, HBT-based amplifiers have relatively low cost, and very low 1/f noise, in addition to the high output and input resistance, which is advantageous in voltage detection applications.

\begin{figure}[ht]
  \centering
  \begin{subfigure}{0.6\textwidth}
  \includegraphics[width=1\linewidth]{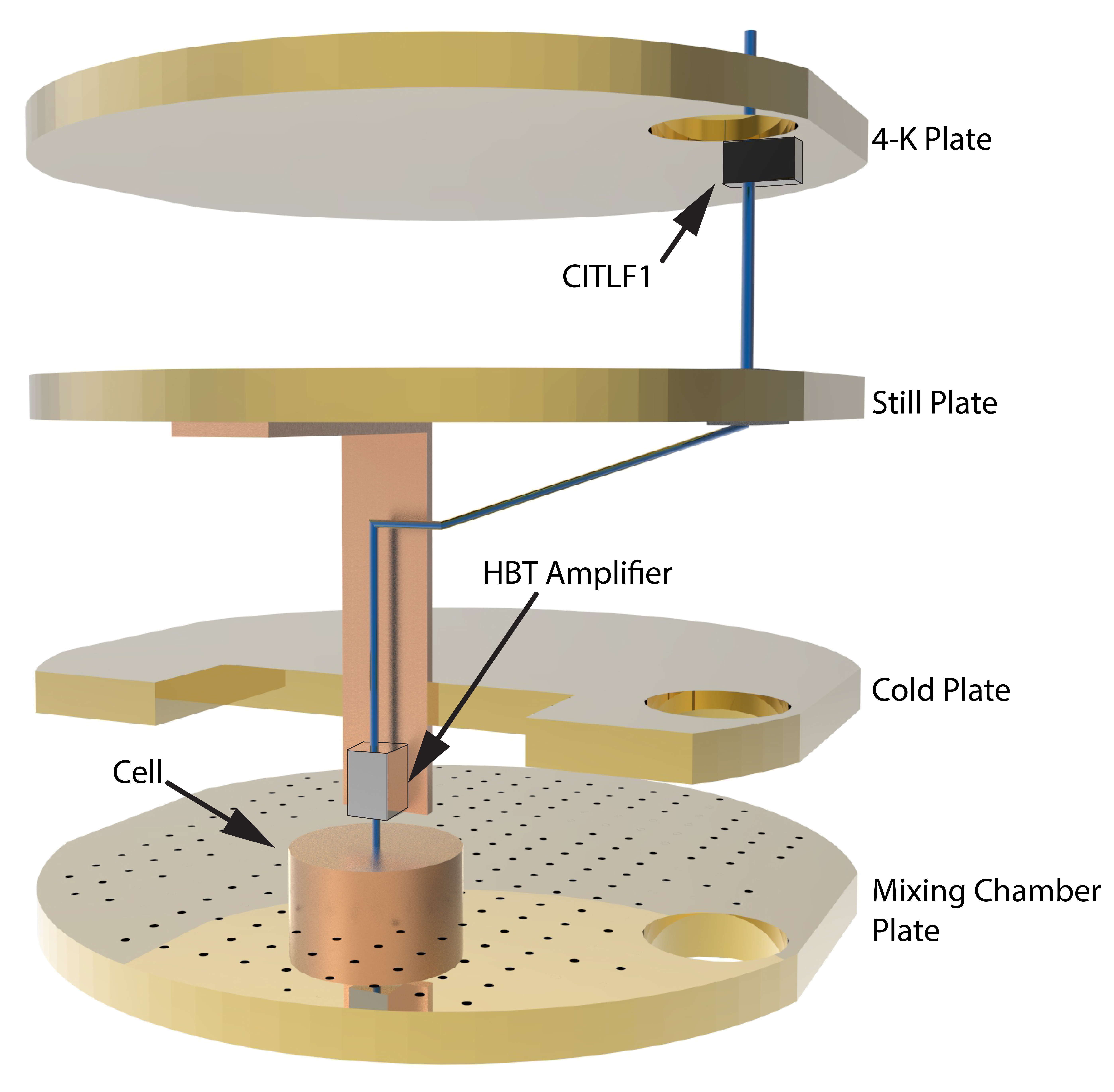}
  \caption{}
  \label{fig:bf}
  \end{subfigure}
  \newline
  \begin{subfigure}{1\textwidth}
  \includegraphics[width=1\linewidth]{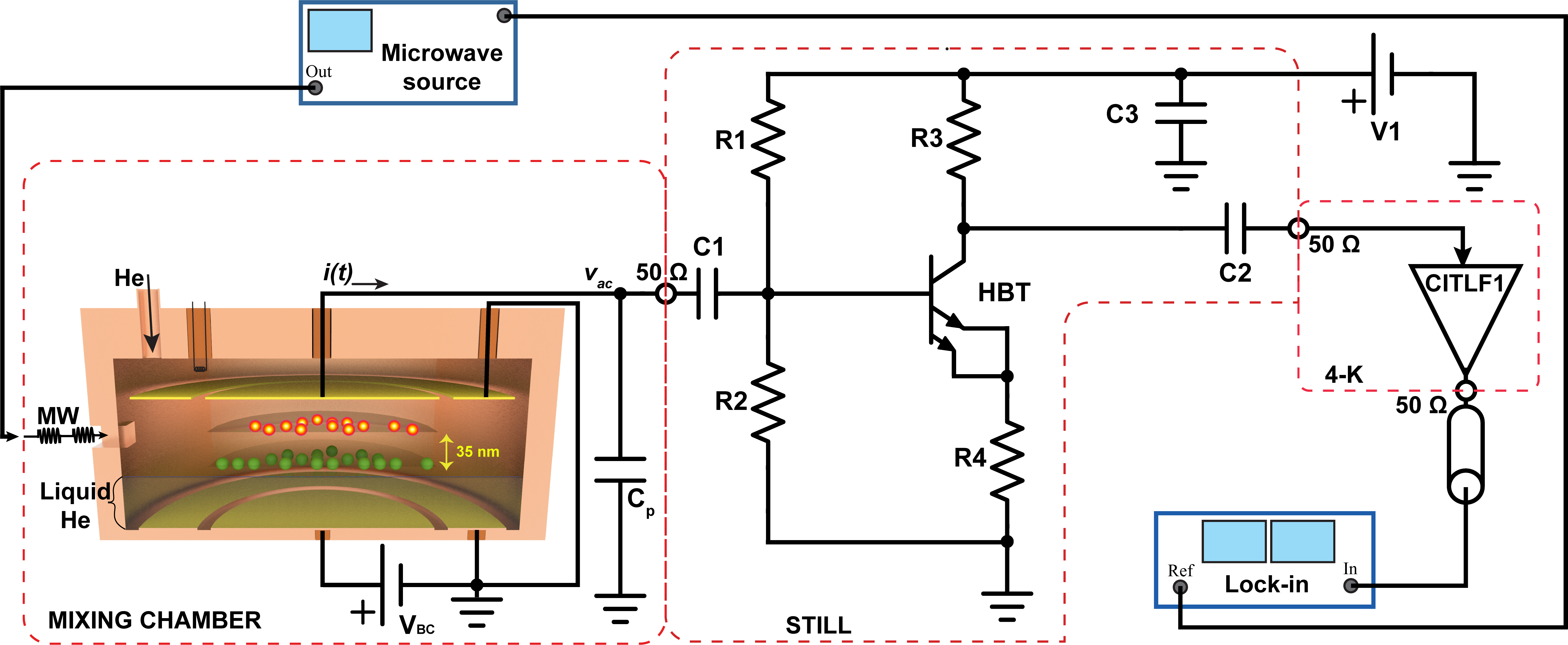}
  \caption{}
  \label{fig:schem}
  \end{subfigure}
  
\caption{\textbf{a} Schematics of the experimental setup for the two-stage amplification showing the first-stage HBT amplifier attached to the still plate by a copper link, and a commercial amplifier (CITLF1) attached to the 4-K plate. \textbf{b} Two-stage amplifier circuit. The image-charge signal $V_{ac}$ due to pulsed MW-excited SE is capacitively coupled to and amplified by a two-stage amplification scheme (HBT+CITLF1). The amplified signal is measured at the room temperature using a conventional lock-in amplifier that is referenced by the modulation frequency of the pulsed MW-excitation.}
\label{fig:Fig1}       
\end{figure}

In this study, we demonstrate an improved method for the image-charge detection of SE on liquid helium ($^3$He) by using a two-stage cryogenic amplification. It consists of a home-made HBT amplifier placed at a short distance from the cell and cooled by attaching it to the still plate of the dilution refrigerator, and a commercially available cryogenic amplifier (Cosmic Microwave Technologies (CMT) CITLF1) cooled by the 4-K plate (see Fig. \ref{fig:bf}). The two-stage amplification was found to provide a \SI{40}{\decibel} constant gain over the wide frequency range [\SI{100}{\kilo\hertz} – \SI{100}{\mega\hertz}] with estimated power dissipation of \SI{90}{\micro\watt} in the first stage. Using this scheme, we demonstrate a significant improvement in the bandwidth and SNR of the detected image-charge signal compared to the previously used method~\cite{Kawakami2019}.

\section{Image-charge detection method}
We start with a brief overview of the image-charge detection method, which was described in detail by Kawakami {\it et al.}~\cite{Kawakami2019}. In the experiment, SE are confined on the surface of liquid helium ($^3$He) placed between two plates of a parallel plate capacitor by applying a positive dc voltage ($V_\textrm{BC}$) to the bottom plate, see the left side of Fig. \ref{fig:schem}. The SE are excited from the ground state to the first excited Rydberg state of the motion perpendicular to the surface of the liquid using microwave (MW) radiation with a fixed frequency $f_{MW}$. To tune the SE into the resonance with MW, their Rydberg transition frequency $f_{21}$ can be adjusted by varying $V_\textrm{BC}$ due to the linear Stark effect. When excited to the first excited state, the SE move away from the surface of liquid helium by a distance $\Delta z \approx \SI{35}{\nano\meter}$. As electrons move closer to the top plate, the change in the image charge induced at the top plate causes an image current $i(t)$ to flow through the capacitor. It is convenient to use a pulsed-modulated MW excitation with the modulation frequency $f_m$ and detect the modulated image-current response by a conventional lock-in amplifier connected directly to the capacitor top plate. The typical value of the RMS image current induced by the pulsed-modulated MW excitation at the modulation frequency $f_m = \SI{100} {\kilo\hertz}$ for SE in a parallel plate capacitor $C_0 = \SI{1}{\pico\farad}$ for SE density $n_s = \SI[retain-unity-mantissa = false]{1e8}{\per\square\centi\meter}$ and the  fractional occupancy of the first excited Rydberg state $\rho_{22}=0.1$ was estimated as $\langle i(t) \rangle = (2\pi f_m e n_s C_0 \Delta z \rho_{22})/\epsilon_0 \approx \SI{100}{\pico\ampere}$, where $\epsilon_0$ is the electrical permittivity of vacuum~\cite{Kawakami2019}. 

The above-estimated current can be detected by a conventional lock-in amplifier, as was demonstrated in Ref.~\cite{Kawakami2019}. However, the available detection BW of commercial lock-in amplifiers is limited to about \SI{5}{\mega\hertz} for the direct current detection, while often it is desirable to increase the detection bandwidth significantly above this value, e.g. for real-time measurements of the image-charge signals. This problem can be solved by using the ac voltage signal instead, which then can be recorded using a high-BW detector.
The ac voltage signal generated is given by $V_{ac}=\Delta q/(C_0+C_p) \approx \SI{10.5}{\nano\volt}$, where $\Delta q$ is the change in the image charge due to the excitation of SE ($=\Delta z e n_s \rho_{22} S / D $), $C_p$ is the parasitic capacitance of cables $\approx \SI{300}{\pico\farad}$, $S/D$ is the ratio between the surface area of a capacitor plate and the distance between the two plates ($S/D \approx \SI{5.65}{\milli\meter)}$. Since the ac voltage signal ($V_{ac})$ generated by direct image-charge  detection is very low, noises make direct detection very difficult. Moreover, the BW is further reduced by the parasitic capacitance of the cables between the top plate and the lock-in amplifier ($ f_c= 1/(2\pi RC_p)$), where $f_c$ is the corner frequency, and $R$ is the cables resistance. Therefore, for higher BW, lowering $C_p$ is important by placing the amplifier in close proximity to the signal source, as well as using high-BW amplifiers. For the above-mentioned reasons, placing a single-stage cryogenic amplifier (ex: CITLF1) at the 4K-plate is not sufficient for high-BW applications.

\section{Two-stage cryogenic amplification}
The image-current measurement setup, which includes our two-stage amplification scheme, is shown in Fig. \ref{fig:schem}. The home-made first-stage HBT amplifier is connected to the capacitor top plate via a short ($\sim \SI{35}{\milli\meter}$) shielded silver-plated cupronickel coaxial cable to reduce the parasitic capacitance ($ C_p \approx \SI{10}{\pico\farad}$, $V_{ac} \approx  \SI{290}{\nano\volt} $) and added noise from the environment. The output of the HBT amplifier is connected to the commercial CITLF1 amplifier located near the 4-K plate via a superconducting NbTi coaxial cable. A stainless steel coaxial cable then connects CITLF1 to a leak-tight SMA (subminiature version A) connector at the room temperature flange of the cryostat, from which the signal is subsequently carried to the lock-in amplifier via a two-meter coaxial cable. The dc bias is applied to both the amplifiers using dc voltage sources located at room temperature.

Few studies have discussed low-noise HBT devices that are capable of operating in the millikelvin range ($< \SI{1}{\kelvin} $) \cite{Curry2015,Wan2017,Curry2019,Zavjalov2019,Ivanov2016,Ivanov2011a,Weinreb2007,Goryachev2010}. Among these devices are Infineon low-noise SiGe HBT RF transistors, which showed improved performance at that temperature range with reduced active noise power dissipation  \cite{Goryachev2010,Ivanov2011a,Ivanov2016,Zavjalov2019}. Infineon NPN SiGe BFP640, in particular, demonstrated enhanced gain and noise-reduction in the millikelvin range without notable degradation \cite{Zavjalov2019,Ivanov2011a,Ivanov2016}. It was indicated that it could achieve low noise temperature ($\sim \SI{2}{\kelvin}$), and a voltage noise as low as $\SI{35}{\pico\volt\per\sqrt{\hertz}}$ at \SI{4.2}{\kelvin} over a frequency range (\SIrange[range-phrase = --]{0.01}{100}{\mega\hertz}) \cite{Ivanov2011a}. Therefore, we have chosen the Infineon HBT transistor at the first stage of our amplification scheme. 

\begin{figure}[ht]
\begin{subfigure}{0.39\textwidth}
  \includegraphics[width=1\linewidth]{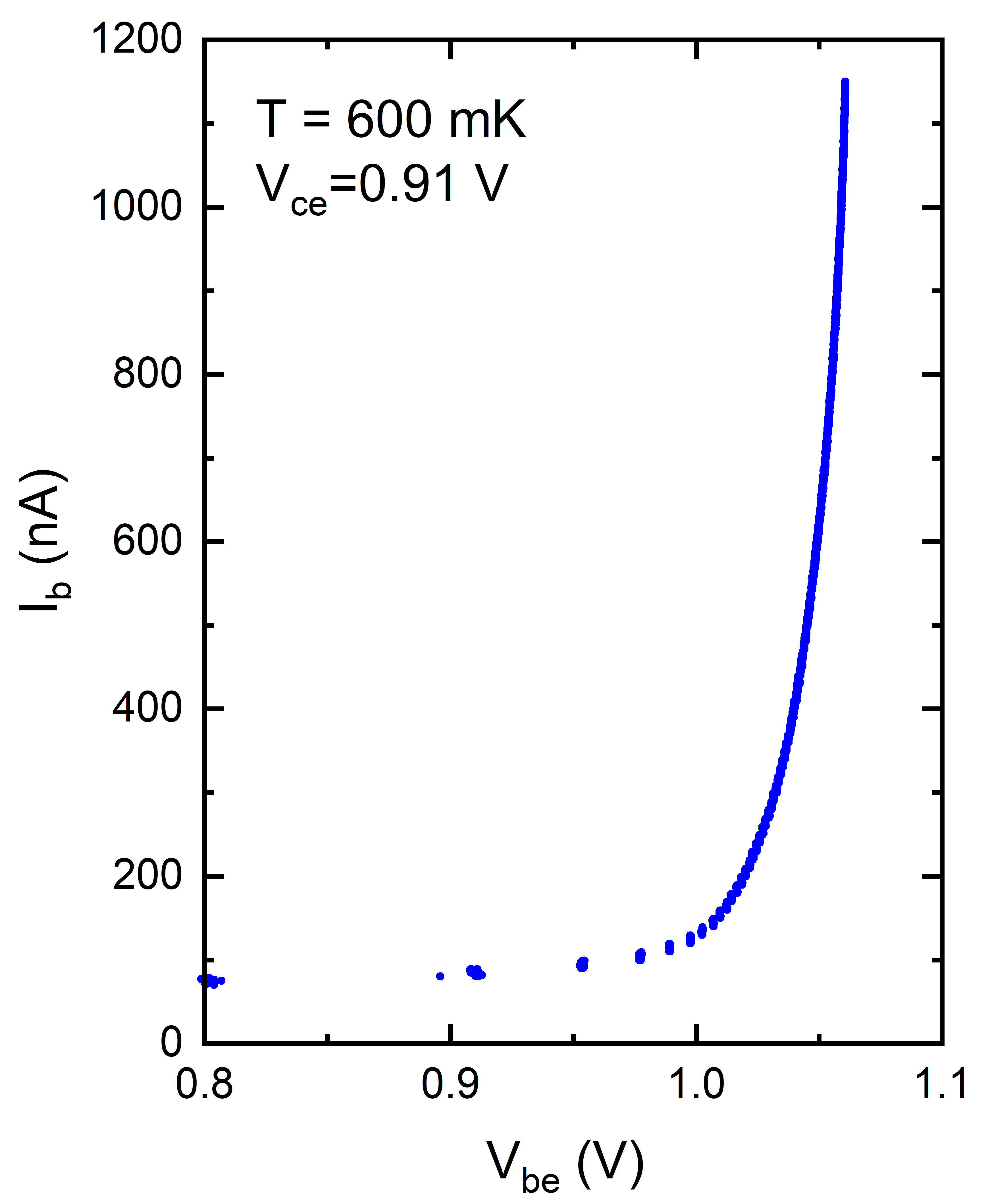}
  \caption{}
  \label{fig:ibvbe}
 \end{subfigure}
 \begin{subfigure}{0.6\textwidth}
 \includegraphics[width=1\linewidth]{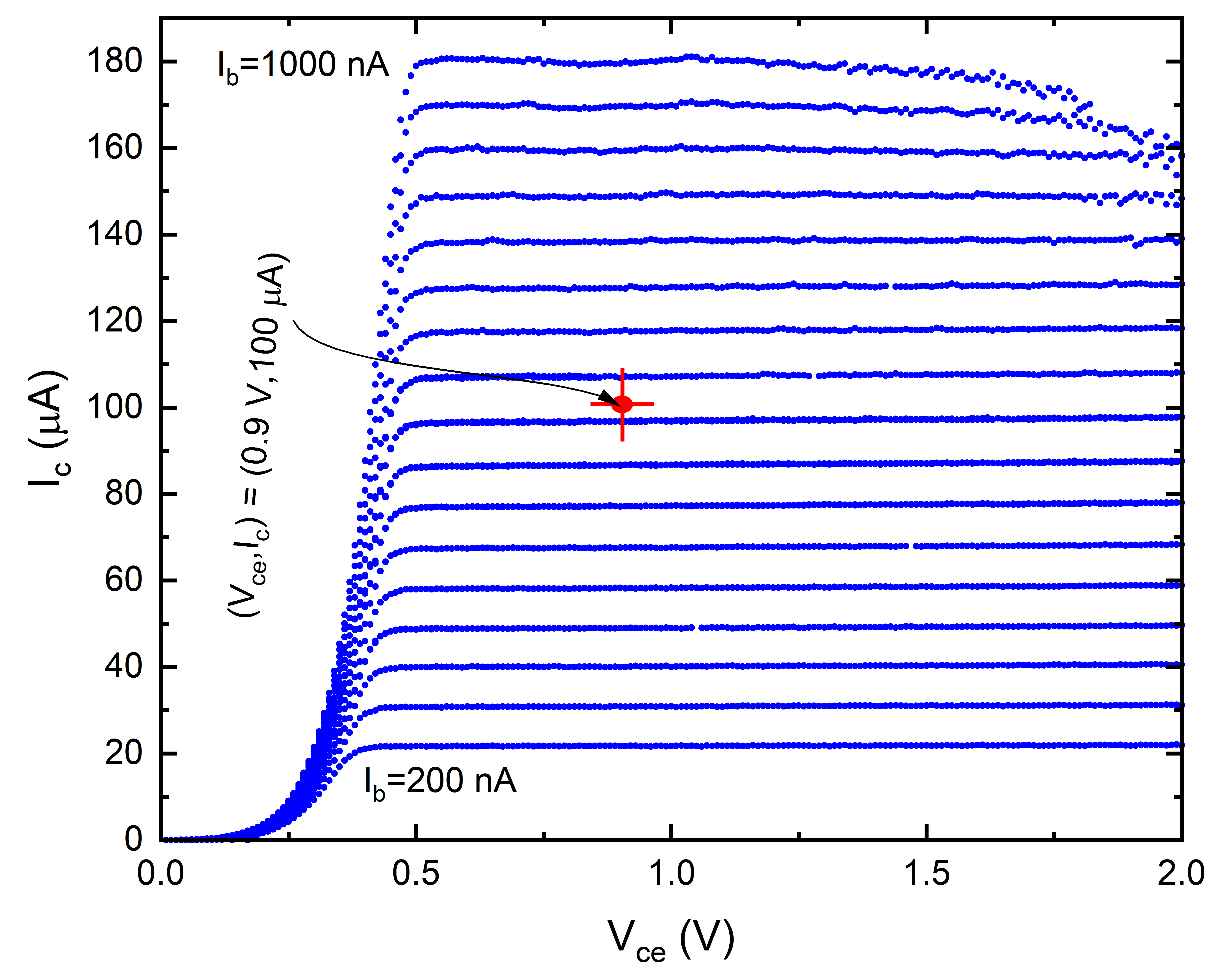}
 \caption{}
 \label{fig:icvce}
 \end{subfigure}
  \newline
  \centering
  \begin{subfigure}{0.6\textwidth}
  \centering
  \includegraphics[width=1\linewidth]{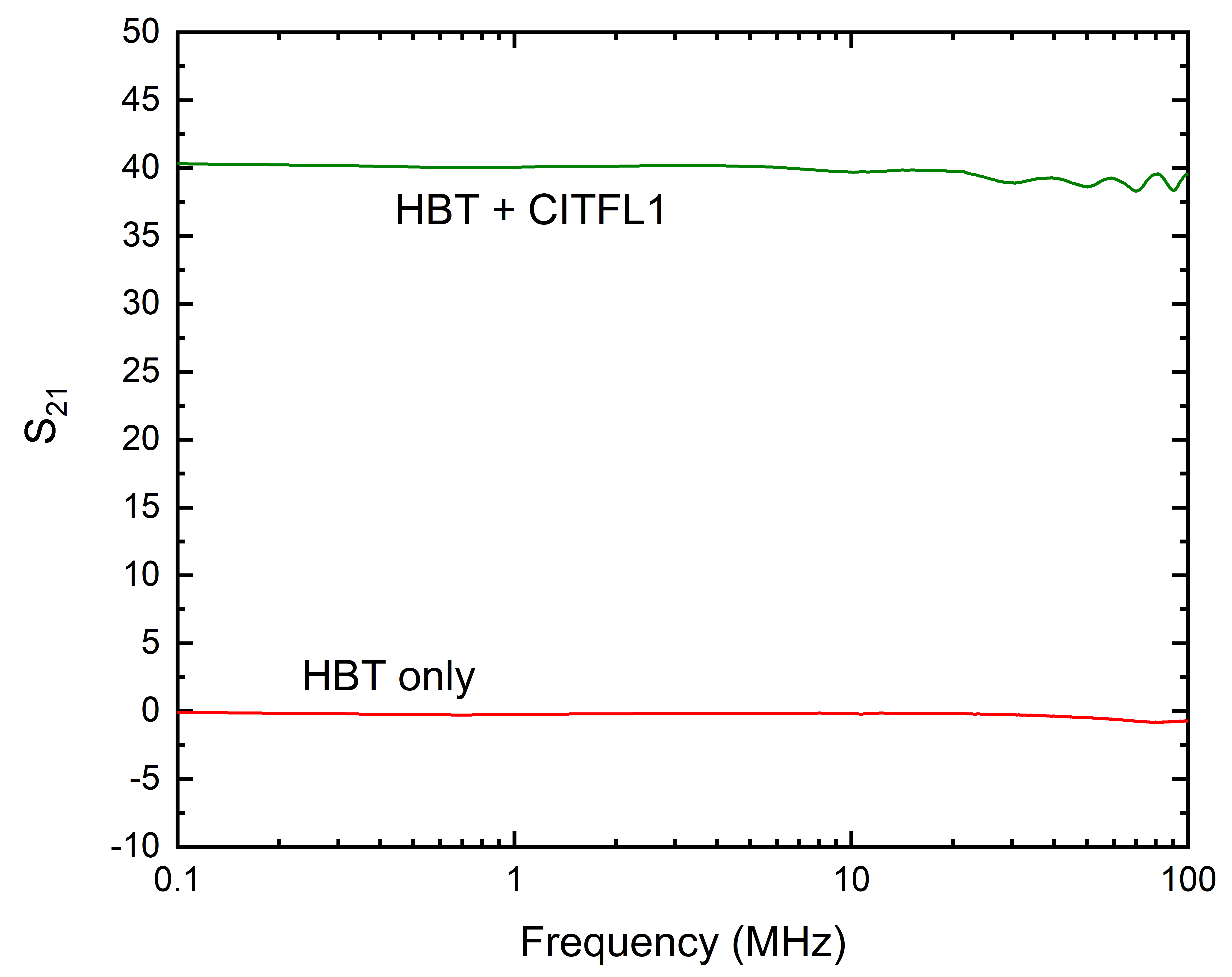}
  \caption{}
  \label{fig:s21}
  \end{subfigure}
\caption{\textbf{a} Input characteristics of HBT measured at $T = \SI{600}{\milli\kelvin}$; the base current $I_b$ versus the base-emitter voltage $V_{be}$ for the collector-emitter voltage $V_{ce} =\SI{0.91}{\volt}$. \textbf{b} Output characteristics of HBT measured at $T = \SI{600}{\milli\kelvin}$; the collector current $I_c$ versus the collector-emitter voltage $V_{ce}$ measured for different values of the base current $I_b$ in the range from \SIrange{200}{1000}{\nano\ampere} separated by $\SI{50}{\nano\ampere}$. The bias operation point is indicated by the red cross. \textbf{c} $S_{21}$-parameter showing the gain of the first stage HBT amplifier (red) and the two-stage (HBT+CITLF1) amplifier (green) versus the signal frequency.}
\label{fig:2}       
\end{figure}

To ensure stability of the amplifier over the target frequency range (\SIrange[range-phrase = --]{0.1}{100}{\mega\hertz}), the design parameters of the amplifier were tuned using a method similar to the one described in Ref.~\cite{Goryachev2010}. It is worth mentioning that not all Infineon HBT transistors with similar type-name exhibit similar characteristics at cryogenic temperatures \cite{Goryachev2010}. It was demonstrated that their low temperature behavior classifies them into three groups and only one of them capable of operating as an amplifier at cryogenic temperatures \cite{Goryachev2010}. Therefore, it is crucial to measure the input and output IV characteristics at low temperatures to examine the transistor before employing it. Figs. \ref{fig:ibvbe} and \ref{fig:icvce} show the input ($I_b(V_{be})$) and output ($I_c(V_{ce})$) characteristics, respectively, of Infineon BFP640ESD transistor~\cite{BFP640ESD}, which was used in our experiment. These data were recorded at $T = \SI{600}{\milli\kelvin}$. The device characteristics show neither hysteresis behavior of $I_c(V_{ce})$ curves nor negative differential resistance, confirming the usability of the device as an amplifier \cite{Goryachev2010}. The $I_c$ curves in Fig.~\ref{fig:icvce} correspond to different values of $I_b$ in the range from 200 to 1000 nA with a step of  \SI{50}{\nano\ampere}.  The curves show almost constant $I_c$ levels at voltages $> \SI{0.5}{\volt}$ in a wide $I_b$ range (\SIrange[range-phrase = --]{200}{800}{\nano\ampere}), which results in an Early voltage (the magnitude of $V_{ce}$ intercept as $I_c(V_{ce})$ characteristics are backward extrapolated  \cite{jaeger})~$V_A \approx \SI{124}{\volt}$, and an intrinsic voltage gain $\mu_f \approx 40V_A \approx 5 \times 10^3$. On the other hand, the intrinsic current gain  $\beta_F  \approx 160$ near the operational bias point ($I_c \approx \SI{0.1}{\milli\ampere}$, $V_{ce} \approx \SI{0.9}{\volt}$) (see red indicator in Fig. \ref{fig:icvce}).

Similar to the previously proposed cryogenic amplifiers \cite{Curry2015,Wan2017,Ivanov2011,Ivanov2016,Goryachev2010}, we use the common-emitter biasing regime with $V_1 = \SI{1}{\volt}$, $R_1 = \SI{235}{\kilo\ohm}$ , $R_2 = \SI{574}{\kilo\ohm}$, $R_3 = \SI{1}{\kilo\ohm}$, $R_4 = \SI{24}{\ohm}$, $C_1 = \SI{12}{\nano\farad}$, $C_2 = \SI{12}{\nano\farad}$, and $C_3 = \SI{220}{\nano\farad}$ (see Fig. \ref{fig:schem}). To ensure reliable operation at the cryogenic temperatures, we used thin-film resistors, as well as NP0 and C0G capacitors in our cryogenic circuit. Moreover, high-frequency SMA and SMP connectors were used to connect the amplifiers to the readout capacitor plate inside the cell. The HBT amplifier is assembled on a lab-built single-layer FR-4 PCB and fixed inside a shielded copper package to provide high thermal conductivity and noise isolation. Additionally, the HBT is covered with a ferrite sheet to provide further noise protection. Although the HBT we used has a power dissipation between (\SIrange[range-phrase = --]{10}{180}{\micro\watt}) at \SI{100}{\milli\kelvin}, we have opted to operate it at a dissipation power of $P \approx \SI{90}{\micro\watt}$ in order to avoid thermal fluctuations and to achieve better stability. To stabilize it further against the thermal fluctuations, the amplifier was placed on the still plate, which has a cooling power of $P \approx \SI{33}{\milli\watt}$ at $T_{MC} \approx \SI{100}{\milli\kelvin}$, instead of the mixing chamber, which has a cooling power of $P \approx \SI{420}{\micro\watt}$ at $T_{MC} \approx \SI{100}{\milli\kelvin}$ in a Bluefors LD400 dilution refrigerator.

The voltage gain of the HBT amplifier, represented by the output-input voltage ratio $S_{21}$ measured at $T = \SI{600}{\milli\kelvin}$, is almost uniform along the target frequency range (\SIrange[range-phrase = --]{0.1}{100}{\mega\hertz}) with a value $\approx \SI{0}{\decibel}$ (unity voltage gain) (see Fig. \ref{fig:s21}). Although higher gain ($A_v \approx \SI{4}{\decibel}$) was achieved with higher bias current, the thermal link between the mixing chamber and the input of the first-stage amplifier through the short cable required operating at lower power. Nevertheless, the main purpose of the first stage of amplification is to improve the overall BW and SNR by reducing the effect of the parasitic capacitance of connecting cables, as well as to provide an impedance matching to maintain the integrity of the signal that is amplified further using the second stage \cite{ashoori1982,Hu2005}. For the second stage, we used the commercial CMT-CITLF1 amplifier which can achieve gain as high as \SI{50}{\decibel} with noise temperature less than \SI{6}{\kelvin} over the frequency range (\SIrange[range-phrase = --]{0.001}{1.5}{\giga\hertz}) \cite{CITLF1}. The gain of the two-stage amplification circuit is shown in Fig. \ref{fig:s21} as $S_{21}$ measured by vector network analyzer (Planar 304/1 VNA). The total gain of the amplifiers is $G \approx \SI{40}{\decibel}$, which is almost constant in the target frequency range (\SIrange[range-phrase = --]{0.1}{100}{\mega\hertz}). This is an important performance characteristic of our device and is required for our experiments.

\section{Detection of the image-charge signal}

\begin{figure}[ht]
\begin{subfigure}{0.505\textwidth}
  \includegraphics[width=1\linewidth]{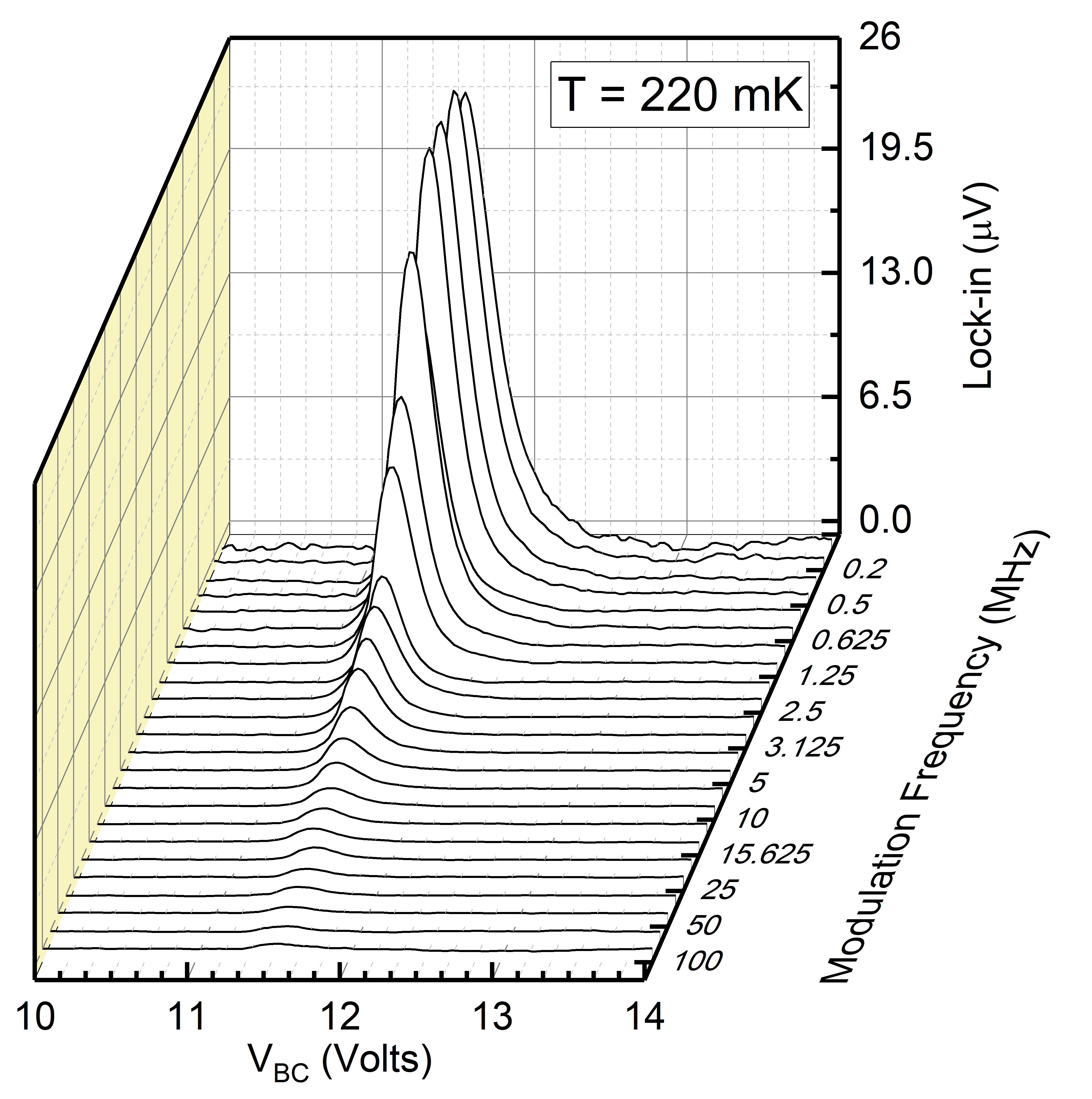}
  \caption{}
  \label{fig:vbcfm-lokin}
  \end{subfigure}
  \begin{subfigure}{0.47\textwidth}
  \includegraphics[width=1\linewidth]{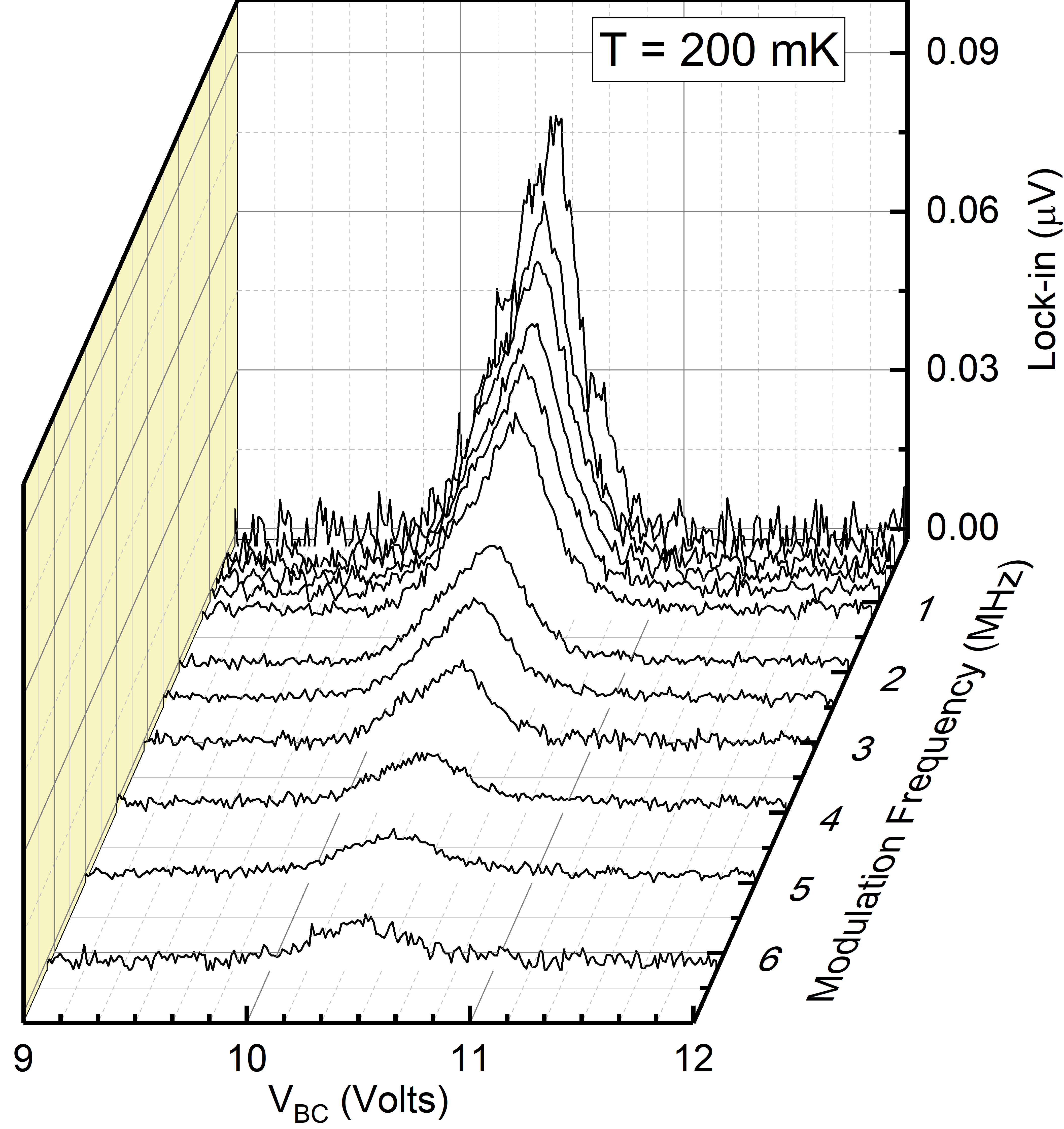}
  \caption{}
  \label{fig:vbcfm-lokin-hbt}
  \end{subfigure}
  \newline
  \begin{subfigure}{1\textwidth}
  \includegraphics[width=1\linewidth]{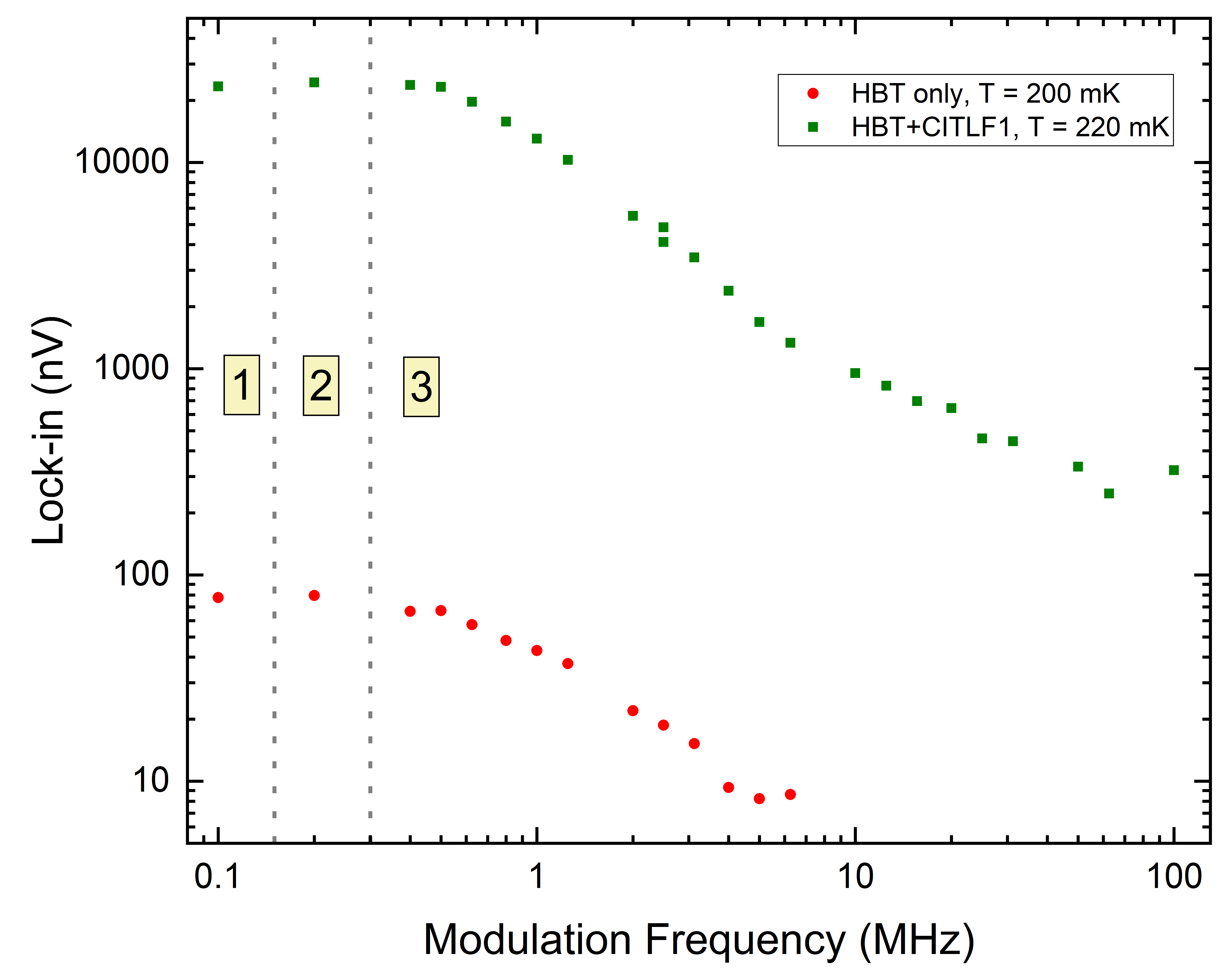}
  \caption{}
  \label{fig:fm-lokin}
  \end{subfigure}
\caption{\textbf{a} Voltage signal amplified by the two-stage amplifiers and measured by the lock-in amplifier for SE at $T=\SI{220}{\milli\kelvin}$ for different modulation frequencies of the applied pulse-modulated MW at $f_{MW}=110$~GHz. \textbf{b} Voltage signal amplified by only the HBT-based amplifier measured by the lock-in amplifier for SE at $T=\SI{200}{\milli\kelvin}$ for different modulation frequencies at similar MW frequency ($f_{MW}$). \textbf{c} Amplitude of the voltage signal detected using the two-stage amplification of SE (green squares: \SI{220}{\milli\kelvin}) in contrast to the signal detected using only the first-stage HBT amplifier (red circles: \SI{200}{\milli\kelvin}). The three marked regions (1, 2, and 3) correspond to three distinct regimes, which determine the maximum detected signal as discussed in the text.}
\label{fig:3}       
\end{figure}

In the experiment described here, the MW radiation at the frequency $f_{MW} = \SI{110} {\giga\hertz}$ is introduced into the cell through a rectangular window via a single-mode waveguide, as shown schematically in Fig.~\ref{fig:schem}. The radiation is pulse (on/off) modulated at a frequency $f_m$, which can be varied in a wide range (\SIrange[range-phrase = --]{0.125}{100}{\mega\hertz}). SE are tuned in resonance with the applied radiation by sweeping the bottom plate voltage $V_\textrm{BC}$, as described earlier. The image-charge signals amplified by the proposed method and recorded by the lock-in amplifier (Stanford Research Systems SR844) for SE at $T_{MC} = \SI{220}{\milli\kelvin}$ and different values of $f_m$ are shown in Fig. \ref{fig:vbcfm-lokin}. An abrupt increase in the voltage signal at $V_\textrm{BC} \approx \SI{11.6}{\volt}$ corresponds to the resonant MW-induced transition of SE from the ground state to the first excited Rydberg state. Fig. \ref{fig:vbcfm-lokin-hbt} shows the image-charge signals amplified by the HBT-based amplifier only at $T_{MC} =\SI{200}{\milli\kelvin}$, where a shift in the resonance condition can be noticed $V_\textrm{BC} \approx \SI{10.45}{\volt}$. This change in peak resonance condition is likely due to the slight difference in the liquid helium level inside the cell, as well as the number of electrons deposited on the helium as the amplifiers setup inside the refrigerator is modified. 

The change in the modulation frequency $f_m$ causes a change in the amplitude of the detected transition signal. Fig.~\ref{fig:fm-lokin} shows the amplitude of the transition signal versus the modulation frequency measured using the two-stage amplification at $T_{MC}=\SI{220}{\milli\kelvin}$ (green squares). For the sake of comparison, the amplitude of the signal measured using only the first-stage HBT amplifier is also shown (red circles), which shows a significant improvement in the signal with the two-stage amplification setup. Three regions are marked to indicate the frequency-dependent amplitude changes, which are classified as: (1) insignificant reduction at $f_m < \SI{150}{\kilo\hertz}$ due to the low corner gain of the amplifiers as specified by the CITLF1 manufacturer \cite{CITLF1}; (2) maximum amplitude regions (\SIrange[range-phrase = --]{150}{300}{\kilo\hertz}); (3) a significant reduction at higher frequencies $f_m > \SI{300}{\kilo\hertz}$ caused by insufficient time for SE to relax from the excited state to the ground state which is theoretically predicted to be on the order of \SI{1}{\micro\second} \cite{Monarkha2007}.

\section{Conclusion}
We improved the image-charge detection of SE on liquid helium by employing voltage detection using a home-made cryogenic low-power HBT amplifier made of commercially available components. In combination with a second-stage commercial amplifier, a wide BW up to \SI{100}{\mega\hertz} and a high gain of \SI{40}{\decibel} were achieved. Using the proposed setup, further high sensitivity measurements can be performed including real-time measurements of the image-charge signals to estimate the time required by SE to relax from the excited Rydberg states to the ground state.  Further improvements to the detected signal might be achieved by using a single-transistor HBT-based cryogenic transimpedance amplifier (TIA), which might demonstrate a superior SNR using current to voltage conversion~\cite{Lin2012a}. 

\begin{acknowledgements}
We acknowledge funding provided by an internal grant from Okinawa Institute of Science Technology (OIST) Graduate University, as well as JST-PRESTO (Grant No. JPMJPR1762) and JSPS KAKENHI (Grant No. 20K15118). We are grateful to V. P. Dvornichenko and A. S. Rybalko for providing technical support.

\end{acknowledgements}

\bibliographystyle{spphys}       

\bibliography{JLTP}

\end{document}